\documentstyle[psfig,12pt]{article}

\def\journal{\topmargin .3in    \oddsidemargin .5in
        \headheight 0pt \headsep 0pt
        \textwidth 5.625in 
\textheight 8.25in 
        \marginparwidth 1.5in
        \parindent 2em
        \parskip .5ex plus .1ex         \jot = 1.5ex}
\journal

\def\ra{\rightarrow}
\begin{document}
\begin{titlepage}
\begin{center}
May 6, 1999      \hfill    LBNL-43248\\
Revised July 6, 1999   \hfill hep-ph/9905478
\vskip .5in

{\large \bf The $Z\ra \overline bb$ decay asymmetry \\
  and flavor changing neutral currents}
\footnote
{This work is supported in part by the Director, Office of Science, Office
of High Energy and Nuclear Physics, Division of High Energy Physics, of the
U.S. Department of Energy under Contract DE-AC03-76SF00098}

\vskip .5in

Michael S. Chanowitz\footnote{Email: chanowitz@lbl.gov}

\vskip .2in

{\em Theoretical Physics Group\\
     Ernest Orlando Lawrence Berkeley National Laboratory\\
     University of California\\
     Berkeley, California 94720}
\end{center}

\vskip .25in

\begin{abstract}

The measured value of $A_{b}$, the $Z \overline bb$ asymmetry parameter, 
disagrees with the Standard Model at 99\% confidence level.
If genuine the discrepancy could indicate new interactions unique to 
third generation quarks, implying enhanced $Z$ penguin amplitudes.  
Enhanced rates are predicted for rare $K$ and $B$ decays, such as 
$K^{+}\ra \pi^{+}\overline \nu\nu$, $K_{L}\ra \pi^{0}\overline 
\nu\nu$, $B\ra X_{s}\overline \nu\nu$, and $B_{s}\ra \overline 
\mu\mu$.  Measurements of $\epsilon^{\prime}/\epsilon$ then imply QCD 
penguin amplitudes must also be similarly enhanced.  The 
Higgs sector of an $SU(2)_L\times SU(2)_R$ gauge theory has some of 
the features needed to explain these phenomena and would also imply 
right-handed penguin amplitudes.

\end{abstract}

\vskip 0.3in
\begin{center}
{\em Note:} The revision includes predictions for additional $K$ and 
$B$ decays.
\end{center}

\end{titlepage}

\renewcommand{\thepage}{\roman{page}}
\setcounter{page}{2}
\mbox{ }

\vskip 1in

\begin{center}
{\bf Disclaimer}
\end{center}

\vskip .2in

\begin{scriptsize}
\begin{quotation}
This document was prepared as an account of work sponsored by the United
States Government. While this document is believed to contain correct
 information, neither the United States Government nor any agency
thereof, nor The Regents of the University of California, nor any of their
employees, makes any warranty, express or implied, or assumes any legal
liability or responsibility for the accuracy, completeness, or usefulness
of any information, apparatus, product, or process disclosed, or represents
that its use would not infringe privately owned rights.  Reference herein
to any specific commercial products process, or service by its trade name,
trademark, manufacturer, or otherwise, does not necessarily constitute or
imply its endorsement, recommendation, or favoring by the United States
Government or any agency thereof, or The Regents of the University of
California.  The views and opinions of authors expressed herein do not
necessarily state or reflect those of the United States Government or any
agency thereof, or The Regents of the University of California.
\end{quotation}
\end{scriptsize}

\vskip 2in

\begin{center}
\begin{small}
{\it Lawrence Berkeley National Laboratory is an equal opportunity employer.}
\end{small}
\end{center}

\newpage

\renewcommand{\thepage}{\arabic{page}}
\setcounter{page}{1}
\noindent {\it \underline {Introduction} } Each successive update of 
the precision electroweak data tends to reinforce the already 
spectacular agreement with the Standard Model (SM).  An exception 
emerged in the Summer 1998 update, when new data from SLC on the $b$ 
quark front-back, left-right polarization asymmetry, $A^{b}_{FBLR}$, 
reinforced a possible discrepancy previously implicit in the LEP 
front-back asymmetry measurement, $A^{b}_{FB}$.  Combined the two 
measurements implied a value for the $b$ asymmetry parameter $A_{b}$ 
three standard deviations ($\sigma$) below the SM value.  The 
discrepancy continues today, though diminished to 2.6$\sigma$ in the 
Spring 1999 data\cite{degroot}, implying inconsistency with the SM at 
99\% confidence level (CL).

The convergence of the SLC and LEP determinations of $A_{b}$ at a 
value in conflict with the SM could resolve the longstanding 
disagreement between the SLC and LEP measurements of the effective 
weak interaction mixing angle, ${\rm sin}^{2}\theta_{W}^{\ell}$, a 
critical parameter that currently provides the most sensitive probe of 
the SM Higgs boson mass.  If $A_{b}$ is affected by new physics then 
$A^{b}_{FB}$ must be removed from the SM fit of ${\rm 
sin}^{2}\theta_{W}^{\ell}$, leaving the remaining measurements in good 
agreement.  This possibility is also consistent with theoretical 
prejudice that the third generation is a likely venue for the 
emergence of new physics.

On the other hand the discrepancy could have an experimental origin.  
The now resolved $R_{b}$ anomaly illustrates the difficulties, which 
may be even greater for $A_{b}$ and $A_{FB}^{b}$.  Or it could be a 
statistical fluctuation.  Unfortunately the study of $Z$ decays is 
nearing its end.  When the dust settles we may still be left wondering 
about the significance of the discrepancy.

The purpose of this paper is to observe that there is another arena in 
which the $A_{b}$ anomaly can be studied.  If it is a genuine sign of 
new physics unique to (or dominant in) the third generation, new 
phenomena must emerge in flavor-changing neutral current (FCNC) 
processes.  Then if the underlying physics has a mass scale much 
greater than $m_{W}$ and $m_{t}$, $Z$ penguin amplitudes are enhanced 
by about a factor two.  The cleanest tests are rare $K$ and $B$ 
decays, such as $K^{+}\ra \pi^{+}\overline \nu\nu$, $K_{L}\ra 
\pi^{0}\overline \nu\nu$, $B\ra X_{s}\overline \nu\nu$, and $B_{s}\ra 
\overline \mu\mu$.  For instance $K^{+}\ra \pi^{+}\overline \nu\nu$ is 
enhanced by a factor 1.9 relative to the SM. A single event has been 
observed, with nominal central value from 1.8 to 2.7 times the SM 
prediction quoted below.\cite{redlinger}

The $A_{b}$ anomaly could arise from new physics in the form of 
radiative corrections or $Z-Z^{\prime}$ or $b-Q$ mixing.  In the first 
case, but not in the latter two, there would generically also be 
enhanced gluon (and photon) penguin amplitudes.  This possibility is 
favored by the recent measurements of $\epsilon^{\prime}/\epsilon$, 
since the $Z$ penguin enhancement by itself exacerbates the 
existing disagreement with the SM, although the theoretical 
uncertainties are considerable. The gluon penguin 
enhancement cannot be deduced in a model independent way 
from the $A_{b}$ anomaly but can be estimated from 
$\epsilon^{\prime}/\epsilon$.  Enhanced $Z$ and gluon 
penguins can be tested in $B$ meson decays and elsewhere.  
They would have a big impact on studies of the CKM matrix and CP 
violation.

\noindent {\it \underline {Fits of the $b$ quark couplings} } In the 
SM the $b$ quark asymmetry parameter is $A_{b}=0.935$ with negligible 
uncertainty.  In terms of the left- and right-handed $Z{\overline bb}$
couplings $g_{bL,R}$ it is
$$
A_{b}={g_{bL}^{2} - g_{bR}^{2} \over g_{bL}^{2} +g_{bR}^{2}}.
\eqno(1)
$$
It is measured directly by the front-back left-right asymmetry, 
$A_{b}= A^{b}_{FBLR} = 0.898 (29)$\cite{degroot} and also by the 
front-back asymmetry using $A_{b}= 4A^{b}_{FB}/3A_{\ell}$ where 
$A_{\ell}$ (${\ell}={\rm e},\mu,\tau$) is the lepton asymmetry 
parameter defined as in eq.  (1) with $b\ra {\ell}$.  Using 
$A^{b}_{FB}= 0.0991 (20)$ from LEP and $A_{\ell}=0.1489 (17)$ from the 
combined leptonic measurements at SLC and LEP, we find $A_{b}= 
0.887(21)$.  The two determinations together imply $A_{b}= 0.891 
(17)$.

I have performed several fits to the five quantities that most 
significantly constrain $g_{bL}$ and $g_{bR}$.  In addition to $A_{b}$ 
and the ratio of partial widths, $R_{b}=\Gamma_{b}/\Gamma_{h}$, they 
are the total $Z$ width $\Gamma_{Z}$, the peak hadronic cross section 
$\sigma_{h}$, and the hadron-lepton ratio $R_{\ell}= 
\Gamma_{h}/\Gamma_{\ell}$.  A brief summary is presented here; details 
will be given elsewhere.\cite{mctobe}

The SM fit assumes ${\rm sin}^{2}\theta_{W}^{\ell}$ = 
0.23128(22) as follows from $A_{\ell}$.
It has chi-squared per degree of freedom $\chi^{2}/dof = 10.4/5$ with 
confidence level $CL= 6.5\%$. In fit 1 $g_{bL}$ and $g_{bR}$ are 
allowed to vary while all other $Z\overline qq$ couplings are held at 
their SM values, yielding $\chi^{2}/dof = 3.0/3$ and 
$CL=39\%$.  In fit 2 only $g_{bR}$ is allowed to vary; the result is
$\chi^{2}/dof = 7.8/4$ with $CL=10\%$, little better than the SM fit. 
In fit 3 the couplings of the $b$, $d$ and $s$ quarks are varied 
equally, $\Delta g_{bL,R}=\Delta g_{dL,R}=\Delta g_{sL,R}$; with a 
result nearly as good as fit 1.  Other fits considered resulted in 
poorer $CL$'s than the SM.

We conclude that positive shifts are preferred for both $g_{bL}$ and 
$g_{bR}$, either for the $b$ quark alone as in fit 1 or for $b$, $d$ 
and $s$ equally as in fit 3.  The need to shift both left and right 
couplings is clear: $\delta A_{b} \simeq -0.05$ requires positive 
shifts in $g_{bR}$ and/or $g_{bL}$ (remember that $g_{bL}<0$) while 
$g_{bL}^{2} + g_{bR}^{2}$ is tightly constrained by the other 
measurements, forcing $\delta g_{bR}^{2} \simeq - \delta g_{bL}^{2}$.  
Fit 3 seems unnatural in that $s,d$ couplings are varied while $u,c$ 
couplings are not, an issue finessed in fit 1 which presumeably  
reflects physics unique to the third generation quarks, perhaps due  
to the large value of the top quark mass.  The 32\% and 5\% contours 
from fit 1 are shown in figure 1, with
the SM values, $g_{bL}, g_{bR}= -0.4197, +0.0771$, and the 
fit central values, $g_{bL}, g_{bR}= -0.4154, +0.0997$.

\noindent {\it \underline {The $Z$ penguin enhancement} } We now focus 
on fit 1 and the FCNC effects it implies.  Physics from higher mass 
scales will couple to the $SU(2)_{L}$ quark eigenstates, so a 
nonuniversal $Z\overline b_{L}b_{L}$ coupling, $\delta g_{bL}$, has 
its origin in a nonuniversal $Z\overline b_{L}^{\prime} b_{L}^{\prime} 
$ amplitude where $b_{L}^{\prime}$ is the weak eigenstate, 
$b_{L}^{\prime} = V_{tb}b_{L} + V_{ts}s_{L} + V_{td}d_{L}$.  As a 
result $Z\overline bs$, $Z\overline bd$, , and $Z\overline sd$ 
interactions are induced.

The very same phenomenon occurs in the SM where the leading 
correction to the $Z\overline bb$ vertex arises from $t$ quark 
loop diagrams. For $m_{t}\ra \infty$ the leading correction 
is\cite{riemann}
$$
 \delta g_{bL}^{\rm SM} =
            {\alpha_{W}(m_{t}) \over 16\pi }
           {m_{t}^{2} \over m_{W}^{2}}     \eqno(2)
$$
where $\alpha_{W} ={\alpha / \rm sin}^{2}\theta_{W}^{\ell}$.  For 
$m_{t}=174.3$ GeV this is $\delta g_{bL}^{\rm SM} \sim 0.0031$.  A 
more complete estimate based on the complete one loop 
result\cite{bernabeu} and with the pole mass $m_{t}$ replaced by the 
running $\overline{MS}$ mass, $\overline m_{t}(m_{t}) \simeq m_{t} - 
8$ GeV,\cite{b-f} yields a similar result, $\delta g_{bL}^{\rm SM} = 
0.0032$, resulting in $g_{bL}=-0.4197$.  In fit 1 $g_{bL}$ is shifted 
by an additional amount, $\delta g_{bL}^{\rm A_{b}} = 0.0043$.   
These are large shifts: e.g., $\delta 
g_{bL}^{\rm SM}$ corresponds to a $3\sigma$ effect in $R_{b}$.

The same Feynman diagrams responsible for the leading $Z\overline bb$ 
vertex correction also generate the SM $Z$ penguin amplitude and in 
the limit $m_{t}\ra \infty$ they are identical.  Rewriting the one 
loop $Z$ penguin vertex for $\overline sd$ transistions as an 
effective $\delta g^{\rm SM}_{\overline sdL}$ coupling  
normalized like $g_{bL}$, we have (see eq.  (2.18) of \cite{b-f})
$$
\delta g^{\rm SM}_{\overline sdL}=\lambda_{t} 
              {\alpha_{W} \over 2\pi}C_{0}(x_{t})
\eqno(3)
$$
where $\lambda_{t}=V^{*}_{ts}V_{td}$, 
$x_{t}=m_{t}^{2}/m_{W}^{2}$, and $C_{0}$ is 
$$
C_{0}(x)=  {x\over 8}\left( {x-6\over x-1} + {3x+2\over (x-1)^{2}}
                       {\rm ln}(x) \right)         \eqno(4)
$$
Taking $m_{t}\gg m_{W}$ and comparing with eq. (2) we have 
$$
\delta g^{\rm SM}_{\overline sdL}=\lambda_{t} \delta g_{bL}^{\rm SM}. 
\eqno(5)
$$

Eq.  (5) shows that if $m_{t}$ were much larger than any other 
relevant scale we could smoothly extrapolate the on-shell $Z\overline 
bb$ vertex correction to the related $Z\overline q^{\prime}q$ 
penguin vertex. The same is true of any new physics at a 
scale $m_{X}\gg m_{W},m_{t}$, whether it affects the $Z\overline bb$ 
vertex by radiative corrections or by $Z-Z^{\prime}$ or $b-Q$ mixing.  
Therefore if the $A_{b}$ anomaly arises from physics at a very high 
scale, the additional contribution to the ${\overline sd}$ $Z$ penguin 
amplitude is
$$
\delta g^{\rm A_{b}}_{\overline sdL}=
       \lambda_{t} \delta g_{bL}^{\rm A_{b}}. 
\eqno(6)
$$
With $\delta g_{bL}^{\rm A_{b}}$ from fit 1, the contribution of 
$\delta g^{\rm A_{b}}_{\overline sdL}$ is equal in sign and magnitude 
to the SM $Z$ penguin, resulting in a factor two enhancement in 
amplitude. 

\noindent {\it \underline {Rare $K$ and $B$ decays}}
The enhancement of the $Z$ penguin implies increased rates for the 
rare $K$ decays $K_{L}\ra \pi^{0}\overline \nu\nu$ and $K^{+}\ra 
\pi^{+}\overline \nu\nu$, and the rare $B$ decays $B\ra X_{s}\overline 
\nu\nu$, and $B_{s}\ra \overline \mu\mu$. The predicted enhancement is 
consistent with the bound on the real part of the $Z$ penguin 
amplitude obtained in \cite{b-s} from $K_{L}\ra 
\overline \mu \mu$.

Predictions for the rare $K$ decays are obtained following \cite{b-s} 
(with parametric updates from \cite{ls}), since $Z_{ds}$ defined in 
\cite{b-s,ls} is $Z_{ds}=\lambda_{t}(C_{0}(x_{t})+C_{b})$ where
$$
C_{b}= {2\pi \over \alpha_{W}}\delta g_{bL}^{A_{b}}  \eqno(7)
$$
with $\alpha_{W}=\alpha/{\rm sin}^{2}\theta_{W}^{\ell}$.  The results 
are
$$
{\rm BR}(K_{L}\ra \pi^{0}\overline \nu\nu)=6.78\cdot 10^{-4}
          ({\rm Im}\lambda_{t})^{2}\left| X_{0}(x_{t}) + C_{b}
          \right|^{2}          
\eqno(8)
$$
and 
$$
{\rm BR}(K^{+}\ra \pi^{+}\overline \nu\nu)=1.55\cdot 10^{-4}
                \left| \lambda_{t}\left( 
                X_{0}(x_{t})+C_{b}\right) + 
                \Delta_{c}\right|^{2}   \eqno(9)
$$
where $\Delta_{c}$ is a nonnegligible charm quark contribution and 
$X_{0}=C_{0}-4B_{0}$ is a combination of the SM $t$ quark $Z$ penguin 
and box amplitudes.  The SM box amplitude, $\sim B_{0}$, is essential 
for gauge invariance and is numerically important in `t Hooft-Feynman 
gauge in which we work.  In the limit $m_{t}\gg m_{W}$ it is 
suppressed by $m_{W}^{2}/m_{t}^{2}$ relative to the penguin because of 
its softer UV behavior.  We assume for the new physics underlying the 
$A_{b}$ anomaly that the penguin amplitude dominates over the box,   
as expected for instance in models with ``hard GIM 
suppression.''\cite{ls}

Similarly, following the parameterization in \cite{b-f}, the $B$ decay 
rates are
$$
{\rm BR}(B\ra X_{s}\overline \nu\nu)= 1.52\cdot 10^{-5}
\left| {V_{ts}\over V_{cb}}\right|^{2}
\left| X_{0}(x_{t})+C_{b}\right|^{2}   \eqno(10)
$$
and 
$$
{\rm BR}(B_{s}\ra \overline \mu\mu)= 3.4\cdot 10^{-9 }\left | Y_{0}(x_{t})
              +C_{b}\right|^{2}   \eqno(11)
$$
where $Y_{0}= C_{0}-B_{0}$.  ${\rm BR}(B\ra X_{d}\overline \nu\nu)$ 
can be obtained by substituting $V_{td}$ for $V_{ts}$ in eq.  (10), 
and ${\rm BR}(B_{d}\ra \overline \mu\mu)$ can be obtained from eq.  
(11) using
$$
{{ \rm BR}(B_{d}\ra \overline \mu\mu)\over {\rm BR}(B_{s}\ra \overline 
\mu\mu)}=   {\tau(B_{d}) \over \tau(B_{s})} {m_{B_{d}}\over m_{B_{s}}}
 {F_{B_{d}}^{2}\over F_{B_{s}}^{2}} 
 {|V_{td}|^{2}\over |V_{ts}|^{2}}.       \eqno(12)
$$

The results are displayed in table 1.  For the $A_{b}$ anomaly the 
branching ratios are enhanced by factors between $\simeq 2$ and 
$\simeq 3$. The SM error estimates are taken from \cite{b-f,b-s,ls}.  
For the $A_{b}$ anomaly two errors are quoted: the first is the 
parametric and theoretical error that is common to the $A_{b}$ anomaly 
and the SM, while the second reflects a $\pm 0.0014$ uncertainty in 
$\delta g_{bL}^{A_{b}}$.  The uncertainties of the ratios are 
dominated by the uncertainty in $\delta g_{bL}^{A_{b}}$ alone.

The ratios differ from unity by about 2.6$\sigma$, which is the 
significance of the $A_{b}$ anomaly itself, whereas the predicted 
branching ratios differ less significantly because of the common 
parametric and theoretical error.  The most significant difference is 
for $B\ra X_{s}\overline \nu\nu$, which has the smallest 
parametric/theoretical error.  Combining all errors in quadrature, the 
predicted SM and $A_{b}$ anomaly branching ratios for $B\ra 
X_{s}\overline \nu\nu$ differ by $2.3 \sigma$.  For $K_{L}\ra 
\pi^{0}\overline \nu\nu$, $K^{+}\ra \pi^{+}\overline \nu\nu$, and 
$B_{s}\ra \overline \mu\mu$ the differences are 1.2$\sigma$, 
1.0$\sigma$, and 1.6$\sigma$ respectively.  The precision of the $K$ 
decay predictions improves as the CKM matrix is measured more 
precisely, while the $B_{s}\ra \overline \mu\mu$ prediction depends on 
the decay constant $F_{B_{s}}$.  If instead of \cite{ls} we take 
$\lambda_{t}$ from the CKM fit of \cite{prs} the precision of the $K$ 
decay predictions is improved.\cite{mctobe}

\noindent {\it \underline {$\epsilon^{\prime}/\epsilon$}} Theoretical 
estimates of $\epsilon^{\prime}/\epsilon$ suggest that the $\delta 
g_{\overline sdL}^{A_{b}}$ $Z$ penguin enhancement is disfavored 
unless QCD penguins are also enhanced.  From the approximate 
analytical formula in \cite{b-s} we find 
$\epsilon^{\prime}/\epsilon=+7.3\cdot 10^{-4}$ for the SM and 
$\epsilon^{\prime}/\epsilon=-0.2 \cdot 10^{-4}$ with the $\delta 
g_{\overline sdL}^{A_{b}}$ $Z$ penguin enhancement.\cite{f1} The most 
recent experimental average (with scaled error) is\cite{sozzi} 
$(\epsilon^{\prime}/\epsilon)_{\rm Expt}= (21.2 \pm 4.6)\cdot 
10^{-4}$.

Taking the theoretical estimates at face value, consistency requires 
that gluon penguins are also enhanced.  If the principal gluon penguin 
term is enhanced by the same factor ($\simeq 2$)
as the $Z$ penguin, the result is $15\cdot 10^{-4}$, while a 
factor 3 enhancement yields $29\cdot 10^{-4}$.

However a large unquantifiable uncertainty hangs over all theoretical 
estimates of $\epsilon^{\prime}/\epsilon$.  Presently they depend 
sensitively on the strange quark running mass and the hadronic matrix 
elements $B_{6}^{1\over 2}$ and $B_{8}^{3\over 2}$, each estimated by 
nonperturbative methods not yet under rigorous control.  Consequently 
we cannot conclude that the SM or the $\delta g_{\overline 
sdL}^{A_{b}}$ enhanced $Z$ penguin are truly inconsistent with the 
data.  The uncertainties will hopefully be resolved by more powerful 
lattice simulations. Until then conclusions based on 
$(\epsilon^{\prime}/\epsilon)$ must be regarded with caution. 

\noindent {\it \underline {Discussion} } The $A_{b}$ anomaly could be 
caused by radiative corrections of new bosons and/or quarks, by 
$Z-Z^{\prime}$ mixing, or by $b-Q$ mixing with heavy quarks $Q$ in 
nonstandard $SU(2)_L$ representations.  Generically radiative 
corrections would also affect gluon and photon penguin amplitudes, by 
model dependent amounts, while $Z-Z^{\prime}$ and $b-Q$ mixing would 
only enhance the $Z$ penguin.  With the major caveat expressed above, 
$\epsilon^{\prime}/\epsilon$ appears to favor radiative corrections, 
since it could be explained if the gluon penguin is enhanced by a 
similar factor to the $\delta g_{\overline sdL}^{A_{b}}$ $Z$ penguin 
enhancement.

The hypothesis that the $A_b$ anomaly represents the effect of higher
energy physics on third generation quarks can be falsified if the 
predicted $Z$ penguin enhancements are absent. If they are present
the hypothesis remains viable and the $A_b$ anomaly provides key
information beyond the FCNC studies. The large value of 
$\delta g_{bR}$ and the condition $\delta(g_{bR}^{2} - g_{bL}^{2}) \gg 
|\delta(g_{bL}^{2} + g_{bR}^{2})|$ would then point to a radical departure 
from the SM with a sharply defined signature. For instance, 
the Higgs sector associated with a right-handed 
extension of the SM gauge sector could shift  
$g_{bR}$ and $g_{bL}$ with little effect on other precision measurements. 
Depending on the right-handed CKM matrix, there could also 
be observable right-handed FCNC effects. 

The burgeoning program 
to study CP violation and the CKM matrix must measure $Z$ and 
gluon penguin amplitudes in order to fully achieve its goals --- an 
enterprise characterized as controlling ``penguin pollution.''  In the 
process we should learn if the FCNC effects implied by fit 1 occur or 
not.  If they do ``penguin pollution'' would be transformed into a 
window on an unanticipated domain of new physics, of which  the 
measurement of $A_{b}$ would have provided the first glimpse. 

\noindent {\it Acknowledgements:} I wish to thank R. Cahn, H. Quinn, 
P. Rowson, S. Sharpe, and especially Y. Grossman and M. Worah for 
helpful discussions.  This work was supported by the Director, Office 
of Science, Office of Basic Energy Services, of the U.S. Department of 
Energy under contract DE-AC03-76SF00098.

\newpage


\begin{figure}
\psfig{file=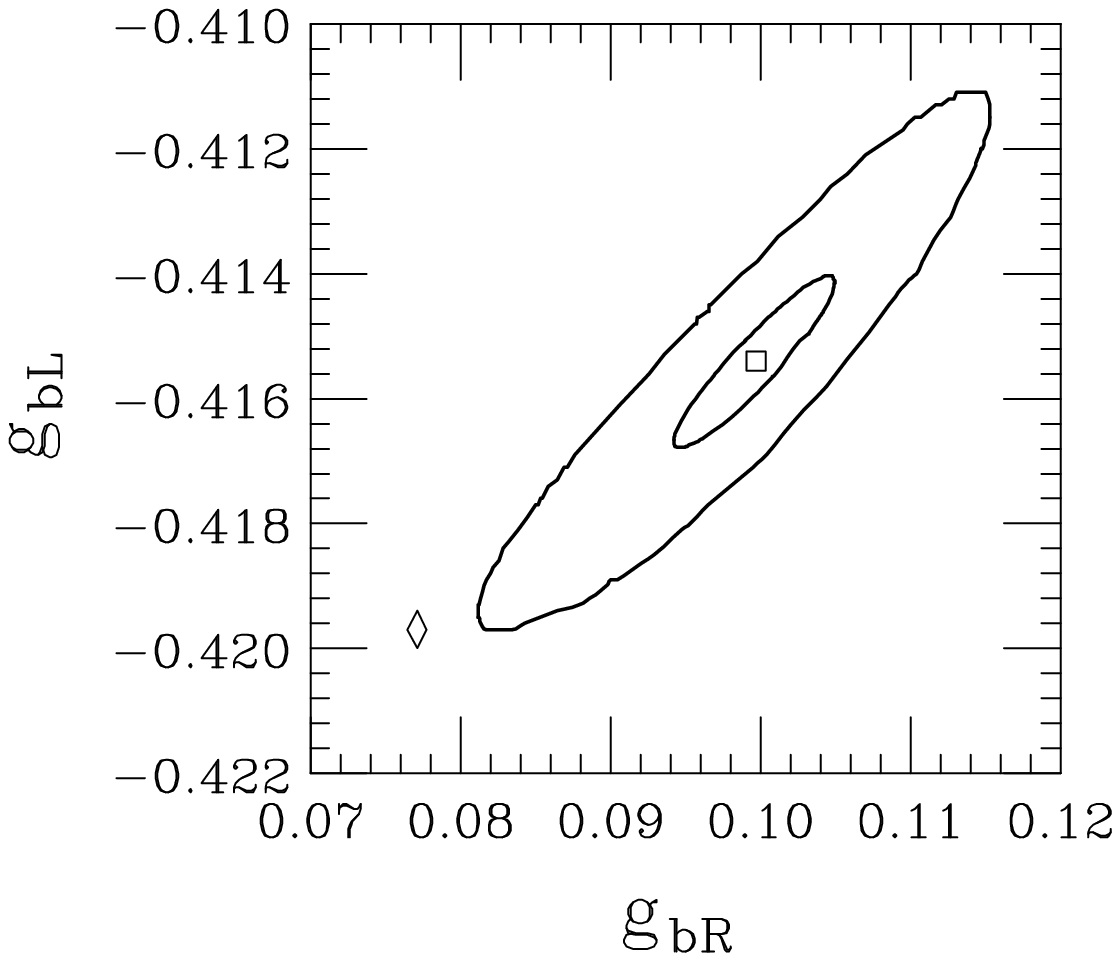,angle=90}
\end{figure}
\noindent Figure 1.  $\chi^{2}$ contours for fit 1. The diamond is the 
SM prediction and the box is the best value from fit 1. The inner 
contour indicates $\chi^{2}=3.5$ corresponding to CL = 32\% for 3 
{\it dof}.  The outer contour indicates $\chi^{2}=7.8$ corresponding 
to CL = 5\% for 3 {\it dof}. 

\vskip .5in

Table 1.  Predicted branching ratios for the $\delta g_{\overline 
sdL}^{A_{b}}$ enhanced $Z$ penguin amplitude and for the SM. The third 
line displays the ratio of the enhanced predictions to the SM.

\begin{center}
\begin{tabular}{c|cccc}
          &$K_{L}\ra \pi^{0}\overline \nu\nu$ 
          &$K^{+}\ra \pi^{+}\overline \nu\nu$ 
          &$B\ra X_{s}\overline \nu\nu$ 
          &$B_{s}\ra \overline \mu\mu$ \\
          &$10^{-11}$ &$10^{-11}$ &$10^{-5}$ &$10^{-9}$\\
\hline
\hline
$\delta g_{\overline sdL}^{A_{b}}$
                          & $6.6\pm 2.4 ^{+1.6}_{-1.4}$
                          & $14.6\pm 5.7 ^{+2.7}_{-2.5}$
                          & $7.8\pm 0.9 ^{+1.9}_{-1.7}$ 
                          & $10.7\pm 3.7 ^{+3.4}_{-2.9}$\\
SM& $2.8\pm 1.1$& $7.7\pm 3.0$& $3.3\pm 0.4$& $3.2\pm 1.1$\\
Ratio& $2.3^{+0.6}_{-0.5}$& $1.9^{+0.4}_{-0.3}$
           &$2.3^{+0.6}_{-0.5}$&$3.3^{+1.1}_{-0.9}$\\
\hline
\hline
\end{tabular}
\end{center}



\begin{thebibliography}{99}
\bibitem{degroot} P.C. Rowson, SLAC-PUB-8132, April, 1999, e-print 
hep-ex/9904016; N. de Groot, presented at {\it Rencontres de 
Moriond}, March 1999, to be published in the proceedings.

\bibitem{redlinger}G. Redlinger (BNL 787 collab.), presented at KAON 
`99, June, 1999, to be published in the proceedings. 

\bibitem{mctobe} M. Chanowitz, to be submitted for publication.

\bibitem{riemann} A.A.~Akhundov, D.Y.~Bardin and T.~Riemann, {\it 
Nucl.  Phys.  B276}:1,1986; W.~Beenakker and W.~Hollik, {\it Z. Phys.  
C40}:141,1988; J.~Bernabeu, A.~Pich and A.~Santamaria, {\it Phys.  
Lett.  200B}:569,1988.

\bibitem{bernabeu}J. Bernabeu, A. Pich, and A. Santamaria, {\it Nucl. 
Phys.}B363:326,1991.

\bibitem{b-f} A.J. Buras and R. Fleischer, TUM-HEP-275-97, to be 
published in {\it Heavy Flavours II}, World Scientific (1997), eds.  
A.J. Buras and M. Lindner, e-Print: hep-ph/9704376.

\bibitem{b-s} A.J. Buras and L. Silvestrini, {\it 
Nucl.Phys.}B546:2991999 and e-Print: hep-ph/9811471.

\bibitem{ls} L. Silvestrini, presented at XXXIV Rencontres de Moriond 
(Les Arc, March, 1999, to be published in the proc., e-Print 
hep-ph/9906202. 

\bibitem{f1} We use $m_{s}=130$ MeV, $B_{6}^{1\over 2}=1.0$ and 
$B_{8}^{3\over 2}=0.8$. 

\bibitem{prs} F. Parodi, P. Roudeau, A. Stocchi, LAL-99-03, Mar 1999, 
e-Print: hep-ex/9903063.

\bibitem{sozzi}M. Sozzi (NA48 collab.), presented at KAON 
`99, June, 1999, to be published in the proceedings. 

\end{thebibliography}
\end{document}